\documentclass[aps,prl,reprint]{revtex4-1}

\usepackage{graphicx}
\usepackage{dcolumn}
\usepackage{bm}

\usepackage{amsmath}
\usepackage[utf8]{inputenc}
\usepackage{amsmath,amssymb,mathtools}
\usepackage{graphicx}
\usepackage[hidelinks, pageanchor=false]{hyperref}
\usepackage{xcolor}
\usepackage{multirow}
\usepackage{babel}
\usepackage{xspace}
\usepackage{enumerate}
\usepackage[justification=Justified]{caption}
\usepackage{subcaption}
\usepackage{orcidlink}
\usepackage{float} 
\usepackage[
  a4paper,
  twoside=false,
  left=1.7cm,
  right=1.7cm,
  top=2.5cm,
  bottom=3cm
]{geometry}

\allowdisplaybreaks
\DeclareRobustCommand{\nn}{\nonumber} 

\def\beq{\begin{equation}}
\def\eeq{\end{equation}}
\def\bea{\begin{align}}
\def\eea{\end{align}}
\def\nn{\nonumber}
\def\Eq#1{Eq.~(\ref{#1})}

\def\ii{\imath 0}
\def\uv{{\rm UV}}

\def\r{{\rm R}}
\def\nn{\nonumber}
\def\lb{\boldsymbol{\ell}}
\def\pb{{\bf p}}

\def\qon#1{q_{#1,0}^{(+)}}

\begin{document}

\preprint{APS/123-QED}

\title{Loop Feynman integration on a quantum computer}

\author{Jorge J. Mart\'{\i}nez de Lejarza\orcidlink{0000-0002-3866-3825}$^1$}\email{jormard@ific.uv.es}
\author{Leandro Cieri\orcidlink{0000-0002-2624-1879}$^1$}\email{leandro.cieri@ific.uv.es}
\author{Michele Grossi\orcidlink{0000-0003-1718-1314}$^2$}\email{michele.grossi@cern.ch}
\author{Sofia Vallecorsa\orcidlink{0000-0002-7003-5765}$^2$}\email{sofia.vallecorsa@cern.ch}
\author{Germán Rodrigo\orcidlink{0000-0003-0451-0529}$^1$}\email{german.rodrigo@csic.es}%
\affiliation{%
 $^1$Instituto de F\'{\i}sica Corpuscular, Universitat de Val\`encia - Consejo Superior de Investigaciones Cient\'{\i}ficas, Parc Cient\'{\i}fic, E-46980 Paterna, Valencia, Spain
}%
\affiliation{%
$^2$European Organization for Nuclear Research (CERN), 1211 Geneva, Switzerland 
}%

\date{\today}

\begin{abstract}

This work investigates in detail the performance and advantages of a new quantum Monte Carlo integrator, dubbed Quantum Fourier Iterative Amplitude Estimation~(QFIAE), to numerically evaluate for the first time loop Feynman integrals in a near-term quantum computer and a quantum simulator. In order to achieve a quadratic speedup, QFIAE introduces a Quantum Neural Network (QNN) that efficiently decomposes the multidimensional integrand into its Fourier series.  For a one-loop tadpole Feynman diagram, we have successfully implemented the quantum algorithm on a real quantum computer and obtained a reasonable agreement with the analytical values. One-loop Feynman diagrams with more external legs have been analyzed in a quantum simulator. These results thoroughly illustrate how our quantum algorithm effectively estimates loop Feynman integrals and the method employed could also find applications in other fields such as finance, artificial intelligence, or other physical sciences.
\end{abstract}

\maketitle
\section{I. Introduction}
\vspace{-0.00cm}
To unravel the mysteries of the universe at its most fundamental level, Quantum Field Theory~(QFT) stands out as an astounding theory, demonstrating significant agreement between theoretical predictions and experimental observations at particle colliders. However, as our comprehension of the particle realm progresses, the need to explore higher energies presents substantial challenges for precise measurements and theoretical predictions. In this regard, the perturbative approach in QFT has gained immense importance, emerging as a crucial framework for deriving accurate predictions in high-energy physics. Nevertheless, computing higher-order contributions in perturbative QFT is far from straightforward. The primary challenge lies in dealing with virtual quantum fluctuations, leading to the intricacies of multiloop-multileg Feynman integrals. The difficulty stems from their multidimensional nature, dependence on multiple scales, and the presence of ultraviolet~(UV), infrared~(IR), and threshold singularities. Tackling these challenges often involves tiresome regularization, renormalization, and subtraction techniques.

Considering the enormous computational demands of standard methods, the imperative to seek alternative approaches becomes evident to overcome existing state-of-the-art limitations. In response, there is a growing interest in exploring innovative strategies rooted in quantum computing to address traditionally challenging problems spanning diverse domains. The potential acceleration offered by quantum computers has spurred various ideas, including Grover's algorithm for efficient database querying~\cite{Grover:1997fa}, Shor's algorithm for factorization of large integers~\cite{shor}, and quantum annealing for Hamiltonian minimization~\cite{qannealing}. In the realm of particle physics, quantum algorithms have made inroads into different areas, such as lattice gauge theories~\cite{Zohar_2016,preskill}. Moreover, these algorithms have been deployed in various tasks related to high-energy colliders~\cite{Delgado:2022tpc}. These applications encompass jet identification and clustering~\cite{thaler,delgado_jets,lejarza,deLejarza:2022vhe}, determination~\cite{carrazza} and integration~\cite{Cruz-Martinez:2023vgs} of parton densities, simulation of parton showers~\cite{spannowsky}, anomaly detection~\cite{Wozniak:2023xbe,Schuhmacher:2023pro, Bermot:2023kvh}, integration of elementary particle processes~\cite{AGLIARDI2022137228}, calculation of color factors in QCD~\cite{Chawdhry:2023jks}, and the establishment of the causal structure of multiloop Feynman diagrams~\cite{Ramirez-Uribe:2021ubp,Clemente:2022nll}. The rapid expansion of applications underscores the versatility of quantum algorithms for a wide array of purposes~\cite{dimeglio2023quantum}.

In light of these recent achievements, this article seeks to investigate the potential of quantum algorithms for the efficient computation of loop Feynman integrals. Specifically, we delve into the capabilities of Quantum Fourier Iterative Amplitude Estimation~(QFIAE), a quantum algorithm recently introduced in~\cite{deLejarza:2023IEEE,tutorial_qfiae}. QFIAE serves as an end-to-end quantum Monte Carlo integrator, showcasing the aimed potential for a quadratic speedup in querying the probability distribution function that follows the target function.

\begin{figure}[t]
       \centering
  \includegraphics[width=0.99\linewidth]{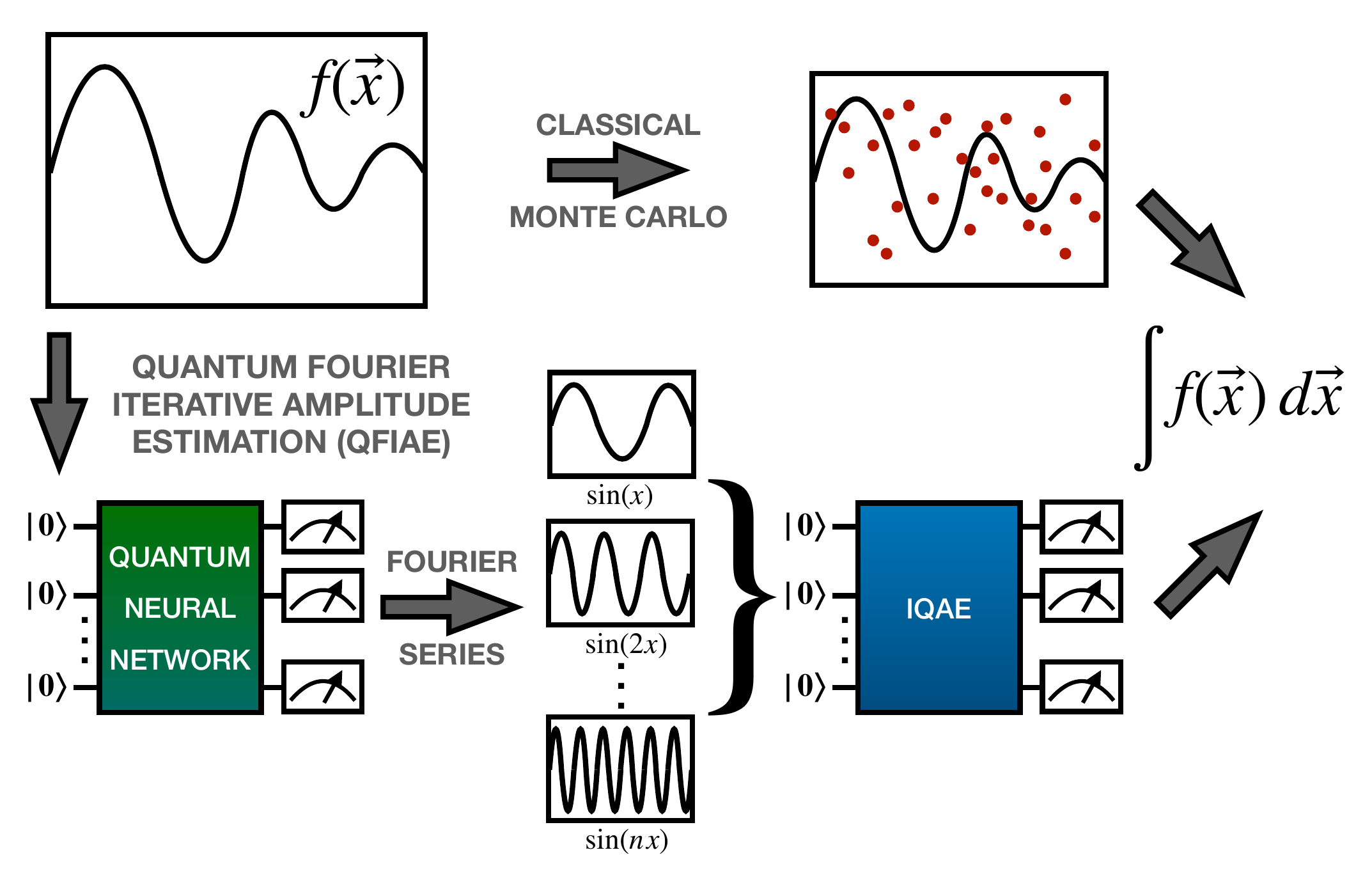}  
        \caption{Comparison of the workflows of classical Monte Carlo integration and the QFIAE quantum algorithm.} 
        \label{fig:qfiae_scheme}
\vspace{-0.5cm}
\end{figure}

The workflow of QFIAE, illustrated in Fig.~\ref{fig:qfiae_scheme}, initiates by decomposing the target function into its Fourier series using a Quantum Neural Network~(QNN) following a data re-uploading approach~\cite{Schuld_2021,Casas:2023ure,P_rez_Salinas_2020}. As was proven in~\cite{Schuld_2021,gil2020input}, following an exponential data encoding approach leads the quantum model to represent a truncated Fourier series. Subsequently, each trigonometric term of the Fourier series undergoes quantum integration using Iterative Quantum Amplitude Estimation~(IQAE)~\cite{Grinko_2021}, an efficient version of Quantum Amplitude Estimation(QAE)~\cite{Brassard_2002}.

The Fourier decomposition enables the target function to be encodable with a minimum number of quantum arithmetic operations. At the same time, it takes advantage of the fact that the sine function is more suitable for integration in a quantum approach. The QNN is the central ingredient of QFIAE that offers a viable strategy to retain the potential quadratic speedup with respect to other quantum integration algorithms recently proposed, such as Fourier Quantum Monte Carlo Integration~(FQMCI)~\cite{Herbert_2022}. The key idea of FQMCI is also to use Fourier series decomposition to approximate the integrand and then estimate each component separately using QAE. Nevertheless, it relies on certain assumptions regarding the acquisition of the Fourier coefficients, which may not hold in general. When these assumptions are not met, the quantum speedup might be wiped out. The QNN ensures a reliable extraction of the coefficients in a quantum way.

The second crucial aspect of QFIAE involves leveraging the advantages of IQAE over QAE. QAE~\cite{Brassard_2002} is a quantum algorithm that estimates quantum state amplitudes using amplitude amplification, a generalization of Grover's algorithm~\cite{Grover:1997fa}. This process enhances the probability of measuring a desired state over a nondesired state. However, QAE has inherent limitations, including its reliance on the resource-intensive Quantum Phase Estimation~(QPE) subroutine~\cite{9781107002173}, which involves operations considered computationally expensive for current Noisy Intermediate Scale Quantum (NISQ) devices that may compromise the anticipated quadratic speedup promised by QAE. To address this challenge, IQAE replaces QPE with a classically efficient post-processing method, reducing the qubits and gates requirements while maintaining the asymptotic quadratic speedup.

\section{II. QUANTUM INTEGRATION OF LOOP FEYNMAN INTEGRALS}
In the realm of particle physics, loop Feynman integrals are mathematical expressions that capture quantum fluctuations arising from virtual particle interactions within subatomic physical processes. These integrals can become highly intricate, often reaching a level of complexity where analytical computation is practically impossible. Consequently, numerical integration methods involving a potentially large number of variables become necessary.

An innovative approach to address such integrals is the Loop-Tree Duality (LTD)~\cite{Aguilera-Verdugo:2020set, Catani:2008xa, Bierenbaum:2010cy, Bierenbaum:2012th}. The LTD methodology transforms loops defined in the Minkowski space of loop four-momenta into trees defined in the Euclidean space of their spatial components. Additionally, it reinterprets virtual states within loops as configurations resembling real-radiation processes.

Among other advantages, this transformation provides a more intuitive understanding of the singular structure of loop integrals~\cite{Buchta:2014dfa,Aguilera-Verdugo:2019kbz}. In particular, the most remarkable property of LTD is the existence of a manifestly causal representation~\cite{Aguilera-Verdugo:2020set,Aguilera-Verdugo:2020kzc,Ramirez-Uribe:2020hes,JesusAguilera-Verdugo:2020fsn,Sborlini:2021owe,TorresBobadilla:2021ivx}, i.e., an integrand representation where certain nonphysical singularities are absent and therefore yields integrands that are numerically more stable. 
Since the integration domain in LTD is Euclidean, not Minkowski, it offers additional advantages, both for analytic applications such as asymptotic expansions~\cite{Driencourt-Mangin:2017gop,Plenter:2020lop}, where the hierarchy of scales is well defined, and numerical applications~\cite{Buchta:2015wna,Driencourt-Mangin:2019yhu} because the number of loop integration variables is independent of the number of external particles. For example, at one loop the number of independent integration variables is always three, the number of spatial components of the loop momentum, although for certain kinematic configurations this number is reduced when the dependence on any of these variables is trivial. This is the case for tadpole and bubble diagrams at one loop. Moreover, LTD offers a unified framework for cross-section calculations, since the dual represen\-tation of loop integrals in Euclidean domains allows a direct combination of virtual and real contributions at the integrand level, resulting in a fully local cancellation of IR~\cite{Hernandez-Pinto:2015ysa,Sborlini:2016gbr,Sborlini:2016hat,Prisco:2020kyb,deJesusAguilera-Verdugo:2021mvg} and UV~\cite{Driencourt-Mangin:2017gop,Driencourt-Mangin:2019aix} singularities without the need for modifying the dimensions of space-time, such as in Dimensional Regularization (DREG)~\cite{Bollini:1972ui,tHooft:1972tcz}.

We now apply QFIAE to this sort of integrals expressed in the LTD formalism, as a proof of concept of how quantum computing has the potential to handle these costly tasks in the particle physics field.
%

\subsection{A. Tadpole loop integral}
The first Feynman loop integral we address is the one-loop tadpole integral 
\beq
{\cal A}^{(1)}_1(m) = \int_{\ell} \frac{1}{\ell^2-m^2+\ii}~,
\eeq
with internal mass~$m$. The $\ii$ factor is the customary complex Feynman prescription for analytic continuation in different kinematical regions. The loop four-momentum to be integrated is~$\ell$. The corresponding mathematical expression in LTD, however, gets support in the loop three-momentum because the loop energy component is integrated out~\cite{Catani:2008xa}. If a local UV counterterm is introduced, then its LTD representation 
\begin{align}
& {\cal A}^{(1,\r)}_1(m;\mu_\uv) \nn \\
& = -\frac{1}{2}\int_{\lb} \left[
\frac{1}{\qon{1}}-\frac{1}{\qon{\uv}}\left(1+ \frac{\mu_\uv^2-m^2}{2 (\qon{\uv})^{2}} \right) \right]~,  
\label{eq:A1}
\end{align}
is well defined in the four physical dimensions, where the integration measure reads $\int_{\lb} = \int d^3\lb/(2\pi)^3$. In \Eq{eq:A1}, we have defined the on-shell energies $\qon{1}=\sqrt{\lb^2+m^2-\ii}$ and $\qon{\uv}=\sqrt{\lb^2+\mu_\uv^2-\ii}$, where $\mu_\uv$ is the renormalization scale. The tadpole integral is one-dimensional because the integrand is independent of the solid angle, and only depends on the modulus of the three-momentum. The change of variable
\beq
|\lb|= \frac{m \,z}{1-z}~, \qquad z=[0,1)~,
\label{eq:changevar}
\eeq
remaps the integration variable into a finite range where the Fourier series of the target function is defined. 

To implement QFIAE in a real quantum computer we must consider that quantum computers operating in the NISQ era encounter diverse sources of noise, ranging from quantum effects like decoherence to hardware-specific errors, including gate, readout, and calibration errors. At the same time, working on the same hardware technology (superconducting qubits), we advocate for a hardware-agnostic implementation strategy. Specifically, we propose an approach where the two modules of QFIAE are implemented on two different quantum computers supplied by different providers. This enables a minimization of the impact of hardware-specific noise on the overall algorithmic performance. 

A one-qubit QNN is trained using an updated version of the Adam gradient descent method first presented in~\cite{Robbiati:2022dkg} and recently improved in~\cite{rtqem}. In this new version of the Adam algorithm, the authors propose a Real-Time Quantum Error Mitigation~(RTQEM) procedure, that allows to mitigate the noise in the QNN parameters during training. We use the full-stack \texttt{Qibo}~\cite{qibo_paper} framework. The high-level algorithm has been written using \texttt{Qibo}, while \texttt{Qibolab}~\cite{qibolab} and \texttt{Qibocal}~\cite{qibocal} are used to respectively control and calibrate the 5-qubit superconducting quantum device hosted in the
Quantum Research Centre (QRC) of the Technology Innovation Institute (TII).

On the other hand, a 5-qubit IQAE algorithm has been executed using \texttt{Qiskit}~\cite{Qiskit} on the 27-qubit IBMQ  superconducting device \textit{ibmq\_mumbai}. To mitigate quantum noise during execution, we employed a pulse-efficient transpilation technique~\cite{Earnest_2021}. Effectively reducing the number of two-qubit gates by harnessing the hardware-native cross-resonance interaction. While it has previously demonstrated promise in the context of Variational Quantum Circuits (VQC)~\cite{Melo2023pulseefficient}, it has not yet been extensively explored for fault-tolerant applications. Considering the positive results presented in Table~\ref{table:hw_tadpole}, this work serves as an intriguing starting point for the applicability of this mitigation technique to fault-tolerant algorithms. Furthermore, we also applied two more error mitigation techniques, Dynamical Decoupling~(DD) and Zero Noise Extrapolation~(ZNE), using the \texttt{Qiskit} Runtime Estimator primitive~\cite{estimator}.  

\begin{table}[t]
\begin{tabular}{cccc}  \hline
Fourier & IQAE & \parbox{3cm}{${\cal A}^{(1,\r)}_1(m;\mu_\uv)$\\
$m=5, \mu_\uv=m/2$} &  
\parbox{3cm}{${\cal A}^{(1,\r)}_1(m;\mu_\uv)$\\
$m=5, \mu_\uv=2m$} \\ \hline
  C  & C & $-0.106$ &  $-0.258$  \\ 
  S  & S & $-0.101(3)$ &  $-0.254(9)$  \\ 
  S  & Q & $-0.108(4)$ &  $-0.270(12)$  \\ 
  Q  & S & $-0.105(2)$ &  $-0.252(6)$  \\
  Q  & Q & $-0.106(3)$ &  $-0.270(9)$  \\ \hline 
  \multicolumn{2}{c}{Analytical}& $-0.1007$ &  $-0.2554$  \\ \hline
\end{tabular}
  \caption{Renormalized tadpole integral ${\cal A}^{(1,\r)}_{1}(m;\mu_\uv)$ on \texttt{Qibo} and IBMQ, as a function of the ratio of the renormalization scale $\mu_\uv$ to the mass $m$, which is fixed to $m=5$~GeV. The Fourier decomposition and IQAE integration are performed either with a quantum simulator (S) or on a real quantum device (Q). In the first raw both components are performed using classical standard methods (C).
  }
  \label{table:hw_tadpole}
\vspace{-0.5cm}
\end{table}
 %

In particular, the hardware implementation has been done by fixing the mass to $m=5$ GeV, and considering two values of the renormalization scale, $\mu_\uv=2m$ and $\mu_\uv=m/2$. 
The uncertainties on the integrals presented in Table~\ref{table:hw_tadpole} have been computed as the quadratic sum of the individual uncertainties provided by IQAE for each trigonometric component. These uncertainties are statistical in nature and are to be combined with the statistical uncertainties from the noise of the quantum devices employed in both the QNN training and IQAE integration modules of the QFIAE algorithm. Furthermore, there are also systematic uncertainties from the classical or quantum Fourier series truncation of the integrand function ${\cal A}^{(1,\r)}_1(m;\mu_\uv)$. Taking into account all statistical and systematic uncertainties the obtained results are in agreement with the analytical values within uncertainties.

The results presented in Table~\ref{table:hw_tadpole} show a relatively small deviation from the analytical value for both $\mu_\uv=m/2$ and $\mu_\uv=2m$. In particular, the agreement with the analytical values when both components of the algorithm are executed on a quantum computer (second to last row of Table~\ref{table:hw_tadpole}) is better than $1.7$ standard deviations in all cases. This result represents a noteworthy achievement for the current state of the art in quantum computing technology. Moreover, considering that it is the first application of an end-to-end quantum algorithm executed on a quantum computer for estimating Feynman loop integrals. Furthermore, part of the deviation comes from the Fourier series approximation itself, as shown in the first row of Table~\ref{table:hw_tadpole}, which constitutes proof of the robustness of the quantum algorithm presented.


\subsection{B. Bubble loop integral}
\begin{figure}[t]
       \centering
  \includegraphics[width=.49\linewidth]{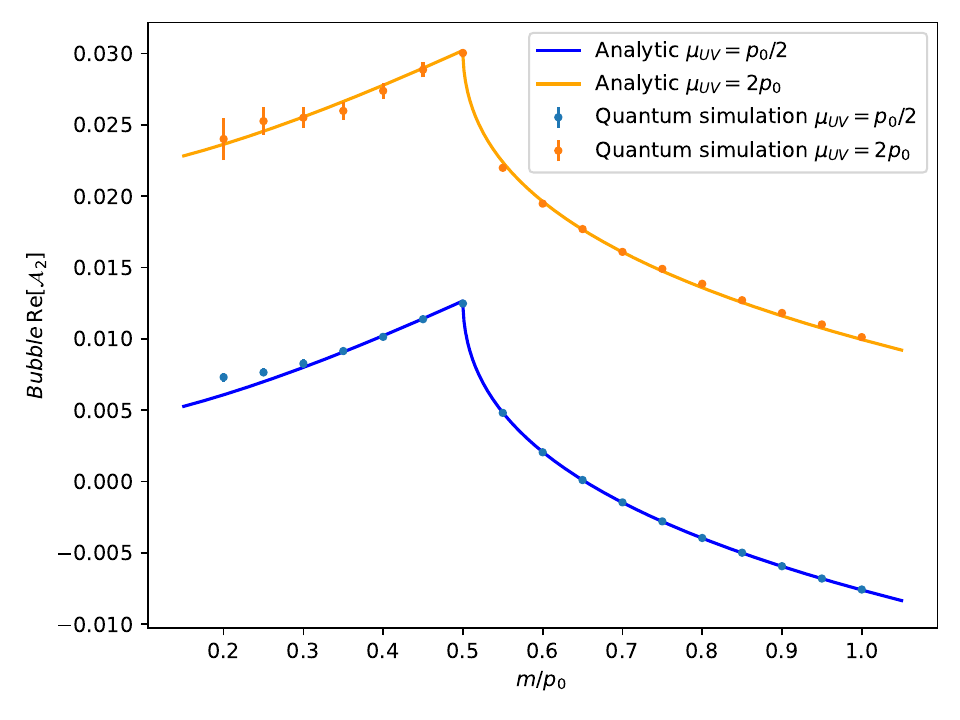}  
  \includegraphics[width=.49\linewidth]{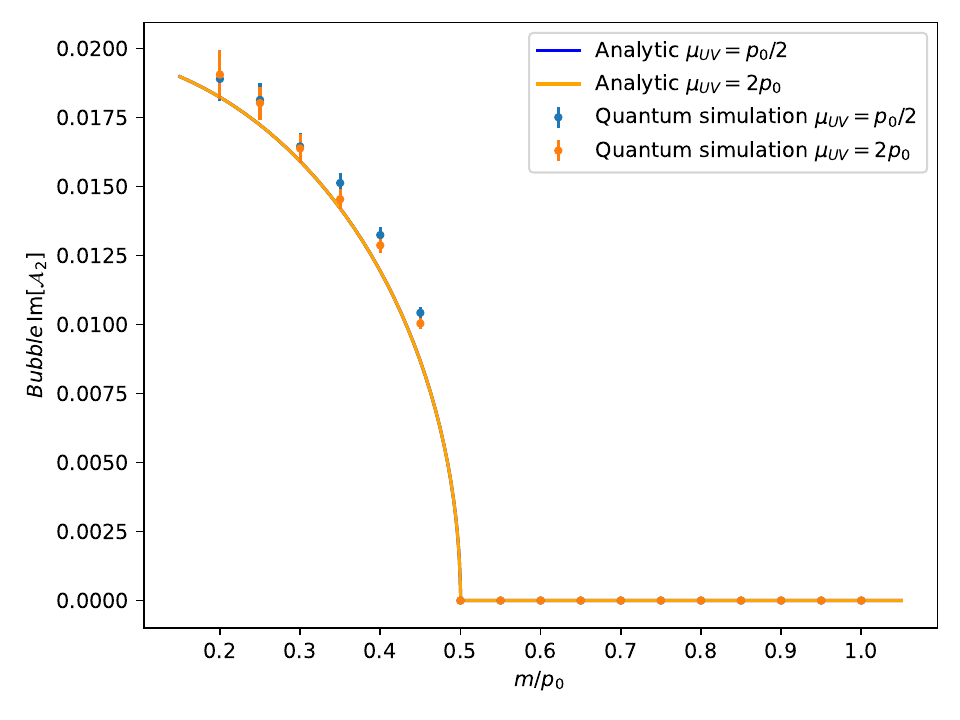}  

        \caption{Quantum integration of the real (left) and imaginary (right) part of the renormalized bubble integral ${\cal A}^{(1,\r)}_{2}(p,m,m;\mu_\uv)$ as a function of the ratio of the mass $m$ to the energy component of the external momentum set at $p_0 = 100$~GeV, and the renormalization scale $\mu_\uv$.} 
        \label{fig:a1bubble}
\vspace{-0.2cm}
\end{figure}

QFIAE is also effective in dealing with Feynman loop integrals with threshold singularities. Therefore, we consider the one-loop bubble integral:
\beq
{\cal A}^{(1)}_2(p,m_1,m_2) =  \int_{\ell} \prod_{i=1}^2
\frac{1}{q_i^2-m_i^2+\ii}~,
\eeq
with $q_1=\ell$ and $q_2=\ell+p$, where $p$ is an external momentum. 
As for the tadpole, we introduce a local UV counterterm to have a finite integral in the UV. The corresponding LTD representation is~\cite{Aguilera-Verdugo:2020set} 
\vspace{-0.1cm}
\beq
{\cal A}^{(1,\r)}_2(p,m_1,m_2) =  \int_{\lb} \left[\frac{1}{x_2} \left( \frac{1}{\lambda^+} 
+ \frac{1}{\lambda^-} \right)-\frac{1}{4(\qon{\uv})^{3}} \right]~,
\eeq
where  $x_2 = \prod_{i=1,2} 2 \qon{i}$, $\lambda^\pm = \sum_{i=1,2} \qon{i} \pm p_0$, and the on-shell energies are given by $\qon{i} = \sqrt{\lb^2+m_i^2-\ii}$ with $i\in \{1,2\}$,
assuming the external momentum has vanishing spatial components, $p=(p_0,{\bf 0})$. If $p_0^2< (m_1+m_2)^2$ the integral is purely real. Otherwise, it gets an 
imaginary contribution from the unitary threshold singularity at $\lambda^-\to 0$, assuming $p_0>0$. To deal with this threshold singularity, it is convenient to introduce a contour deformation in the complex plane, in order to smooth the behavior of the function in the vicinity of the threshold without altering the result of the integral. We consider $m_1=m_2=m$, and proceed to estimate the integral using two different QNNs to fit separately the real and imaginary parts of the integrand and then integrate each Fourier series using IQAE. 
We have implemented QFIAE on two different quantum simulators. For the QNNs, we employed \texttt{Pennylane}~\cite{pennylane} whereas for IQAE, we utilized \texttt{Qibo}.\\
\hspace*{2mm} Results are depicted in Fig.~\ref{fig:a1bubble}, where the real and imaginary parts are displayed on the left and right plots, respectively.  
The statistical uncertainties are calculated as in the previous section. It is worth mentioning that uncertainties in the region below the unitary threshold ($m/p_0>0.5$) 
appear to be larger compared to the rest. This is explained by the Fourier coefficients for low frequencies being larger than in the high-mass region. Hence the statistical uncertainties on the integrals of the trigonometric functions for low frequencies are intrinsically larger since the integrals we are estimating are also larger, i.e. have a larger weight, which is the Fourier coefficient. Another interesting point about these results is that in the low-mass region, the QNNs fit the target function with a slight drop in performance, hence there seems to be a correlation between the QNN struggling to fit a function and the coefficients of the lower frequency terms being larger.\\
Despite this, it is remarkable that the quantum integration values are in agreement with the analytical values within uncertainties. This constitutes another significant achievement since we have successfully circumvented the threshold singularity while applying a quantum algorithm to estimate the bubble integral. 

\subsection{C. Triangle loop integral}
We now consider the one-loop three-point function, which corresponds to a triangle loop topology.
The LTD representation of this integral is given by~\cite{Aguilera-Verdugo:2020kzc}
\beq
\begin{split}
{\cal A}^{(1)}_3&(p_1,p_2,m_1, m_2,m_3) = - \int_{\lb} \frac{1}{x_3}
\left( 
\frac{1}{\lambda_{12}^- \lambda_{23}^+} \right. \\
&\left.+ \frac{1}{\lambda_{23}^+ \lambda_{31}^-}  
+  \frac{1}{\lambda_{31}^- \lambda_{12}^+} 
+ (\lambda_{ij}^+ \leftrightarrow \lambda_{ij}^-) \right)~,
\end{split}
\label{eq:triangle}
\eeq
with $x_3 = \prod_{i=1}^3 2 \qon{i}$, where now the on-shell energies are $\qon{1}=\sqrt{(\lb+\pb_1)^2+m_1^2-\ii}$, and $\qon{i} = \sqrt{\lb^2+m_i^2-\ii}$ for $i\in\{2,3\}$. We work in the center of mass frame where $\pb_{12}=\pb_1+\pb_2=0$, $p_{12,0}= p_{1,0}+$ $p_{2,0}=\sqrt{s}$ and the external momenta $p_1$ and $p_2$ are back-to-back along the $z$ axis. The causal denominators are
\beq
\begin{split}
\lambda_{31}^\pm &= \qon{3} + \qon{1} \pm p_{1,0},~ 
\lambda_{12}^\pm = \qon{1} + \qon{2} \pm p_{2,0},\\ 
  \lambda_{23}^\pm &= \qon{2} + \qon{3} \mp p_{12,0}.
\end{split}
\eeq

\begin{figure}[t]
       \centering
  \includegraphics[width=.49\linewidth]{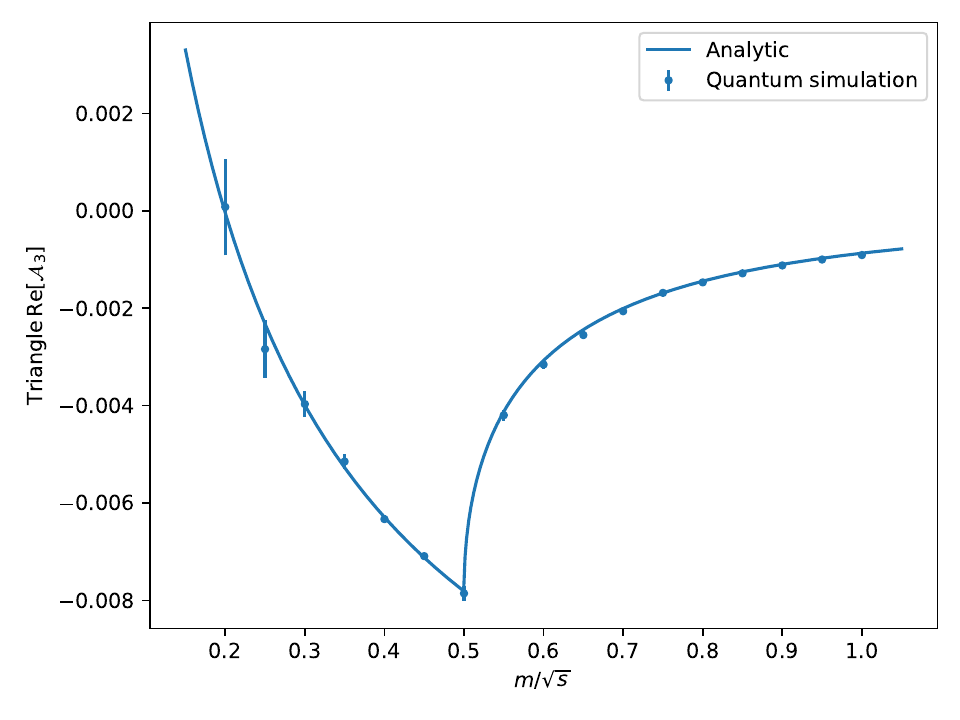}  
  \includegraphics[width=.49\linewidth]{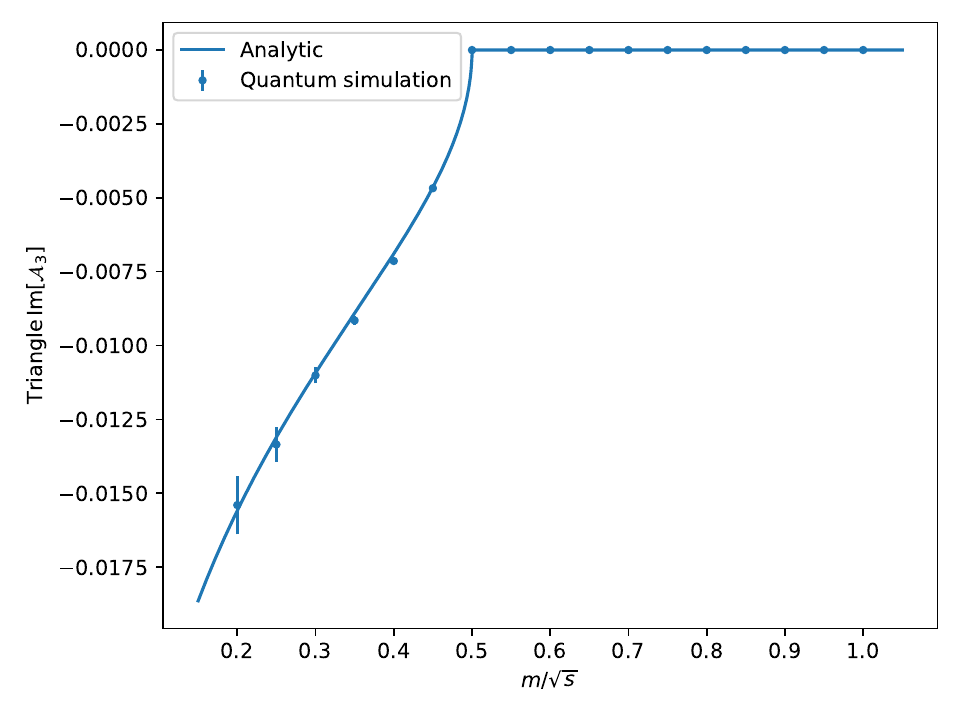}  
        \caption{Quantum integration of the real (left) and imaginary (right) part of the triangle integral ${\cal A}^{(1)}_{3}(p_1,p_2,m,m,m)$ as a function of the ratio of the mass $m$ to the cms energy set at $\sqrt{s}= 2$~GeV.
        }
        \label{fig:a1triangle}
\end{figure}

The integration variables in~\Eq{eq:triangle} are the modulus of the loop three-momentum and its polar angle with respect to $p_1$, considering that the azimuthal integration is trivial. This means that the Fourier decomposition is a function of two variables and we have to integrate each of them separately. 
We also apply a contour deformation to deal with the unitary threshold singularity at $\lambda_{23}^+ \to 0$, when $s>(m_2+m_3)^2$. The contour deformation smooths the behavior of the integrand over the threshold singularity and therefore significantly improves the quality of the Fourier decomposition. 
The estimations of the triangle integral, obtained in \texttt{Pennylane} and \texttt{Qibo} simulators, are shown in Figs.~\ref{fig:a1triangle}(left) and~\ref{fig:a1triangle}(right), illustrating the real and imaginary components, respectively. The statistical uncertainties are calculated as in the previous sections and are expected to be higher since we are performing a double IQAE integral, and each integration introduces an error. However, the estimated uncertainties align well with the deviations observed in the real and imaginary components of the integral. All in all, this represents another noteworthy accomplishment as we have successfully extended for the first time the QFIAE algorithm to a two-dimensional function with a threshold singularity, and utilized a quantum algorithm to approximate the integral with sufficient accuracy. 
\section{III. CONCLUSIONS}
In this work, we have for the first time successfully implemented a quantum algorithm for evaluating Feynman loop integrals in quantum hardware.
Our approach is based on the nonconventional LTD representation of Feynman loop integrals and the QFIAE quantum algorithm which provides a theoretical quadratic quantum speedup in the number of queries of the probability distribution function. 

We have provided a clear indication of how the optimization of a quantum circuit, at the level of pulses and the application of combined error mitigation techniques, enable the collection of accurate results. Specifically, we employed QFIAE to numerically integrate the one-dimensional tadpole Feynman loop integral entirely on real quantum hardware provided by \texttt{Qibo} and IBM. Moreover, loop Feynman integrals with more external legs and dimensions have also been integrated using QFIAE on quantum simulators across various kinematical regions, circumventing different threshold singularities using a contour deformation. The obtained results demonstrate the remarkable performance of QFIAE in estimating these integrals. 

Overall, our study highlights the potential of quantum Monte Carlo integrators, particularly the QFIAE quantum algorithm, in integrating Feynman loop integrals. These findings, some of them obtained from quantum superconducting devices, serve as an initial step toward the search for a quantum method capable of speeding up the computationally expensive task of evaluating Feynman multiloop-multileg integrals. Moreover, they offer a proof of concept for the method's applicability to wider challenges across various domains that require Monte Carlo integration, such as finance, artificial intelligence and other physical sciences.

When considering the scalability of QFIAE to higher dimensions, it is important to analyze its two key components separately. Regarding the IQAE integration, the use of Fourier decomposition allows us to express a multidimensional integral as a product of individual integrals containing Fourier terms with independent integration variables. Consequently, regardless of the dimensionality, IQAE is always applied to one-dimensional trigonometric integrands after Fourier decomposition. This ensures its efficient extension to multidimensional functions. In relation to the QNN fitting the target function, universality has been theoretically demonstrated in \cite{Schuld_2021} for exponential encoding and a data re-uploading scheme. This constitutes a theoretical proof that the first component of QFIAE is also scalable to higher dimensions. 

Still, the main challenge lies in
identifying a suitable Ansatz capable of ensuring the trainability of multidimensional functions, which is essential for addressing higher-loop integrals. While universality has proven effective in lower dimensions, extending it to higher dimensions remains an open problem. In this work, we achieved a successful fit for one and two-dimensional functions. However, the search for a viable Ansatz in higher dimensions which involves a tradeoff between trainability and expressivity of the variational quantum circuit is left for future work.

Progressing in the previous limitation would assure, according to the algorithm proposed in this article, a significant noteworthy advancement over classical Monte Carlo methods for computing these integrals, presenting new opportunities for unlocking previously inaccessible levels of precision in theoretical predictions for high-energy physics at colliders.

\vspace{-0.00cm}
\section{Acknowledgments}
\begin{acknowledgments}
We acknowledge Matteo Robbiati for providing help and support in executing the QNN on the \texttt{Qibo} hardware and Stefano Carrazza for providing us access to the quantum device. We acknowledge financial support from the Spanish Government (Agencia Estatal de Investigaci\'on MCIN/AEI/ 10.13039/501100011033) Grant No. PID2020-114473GB-I00, and Generalitat Valenciana Grants No. PROMETEO/2021/071 and ASFAE/2022/009 (Planes Complementarios de I+D+i, NextGenerationEU).  
This work is also supported by the Ministry of Economic Affairs and Digital Transformation of the Spanish Government and NextGenerationEU through the Quantum Spain~project, and by CSIC Interdisciplinary Thematic Platform on Quantum Technologies (PTI-QTEP+).
JML is supported by Generalitat Valenciana (Grant No. ACIF/2021/219) and CSIC (Grant No. IMOVE23124). LC is supported by Generalitat Valenciana GenT Excellence Programme (CIDE\-GENT/2020/011) and ILINK22045. MG and SV are supported by CERN through the CERN QTI. Access to the IBM Quantum Services was obtained through the IBM Quantum Hub at CERN.
\end{acknowledgments}

\bibliographystyle{apsrev4-1}
\bibliography{bibliography}


\onecolumngrid
\clearpage
\newpage
\subsection{Appendix A: Hardware implementation of QFIAE}
\label{subsection:qfiae_hw}

Although the implementation of Monte Carlo integration in quantum simulators is of great interest for proof of concept purposes, the potential quantum advantage will only be materialized when the quantum algorithm is run on a real quantum device. To this aim, we have addressed the challenge of implementing an end-to-end quantum integrator method into two different quantum devices.

First, the QNN has been trained using an updated version of the Adam gradient descent method first presented in \cite{Robbiati:2022dkg} and recently improved in \cite{rtqem}. In this new version of the algorithm, the authors propose a Real-Time Quantum Error Mitigation (RTQEM) procedure, that allows to mitigate the noise in the QNN parameters during training.
We use the \texttt{Qibo} \cite{qibo_paper} framework to compute the full-stack procedure. The high-level quantum computing algorithm has been written using \texttt{Qibo}, while \texttt{Qibolab} \cite{qibolab} and \texttt{Qibocal} \cite{qibocal} are used to respectively control and calibrate the 5-qubit superconducting device employed.

In this case, a more hardware-friendly linear Ansatz has been chosen to construct the QNN in one qubit. In particular each layer $\mathcal{L}_{LA}^{(l)}(\vec{x},\vec{\theta})$ is defined as:
\begin{equation}
    \mathcal{L}_{LA}^{(l)}(\vec{x},\vec{\theta})=\prod_{i=1}^MR_z(\theta_3 x_i+\kappa\theta_4)R_y(\theta_1 x_i+\theta_2), \quad \textrm{with} 
        \begin{cases}
        \kappa=1 & \text{if } l \text{ is the last layer,} \\
        \kappa=0 & \text{otherwise.}
    \end{cases}
    \label{eq:linear_ansatz_hw}
\end{equation}

On the other hand, the IQAE has been executed using \texttt{Qiskit} \cite{Qiskit} on the IBMQ 27-qubits device \textit{ibmq\_mumbai}. The quantum circuits of the operators $\mathcal{A}$ and $\mathcal{Q}$ of the IQAE algorithm are presented in Fig. \ref{fig:qaecircuits}. To mitigate quantum noise during the execution of this algorithm, we employed a pulse-efficient transpilation technique \cite{Earnest_2021}. This technique effectively reduces the number of two-qubit gate operations by harnessing the hardware-native cross-resonance interaction, potentially leading to a reduction in quantum noise. While it has previously demonstrated promise in the context of Variational Quantum Circuits (VQC) \cite{Melo2023pulseefficient}, it has not yet been extensively explored for fault-tolerant applications. Hence, this work serves as an intriguing starting point for the applicability of this mitigation technique to fault-tolerant algorithms, given the positive results presented in Fig. 3. Furthermore, we also applied two more error mitigation techniques, Dynamical Decoupling (DD) and Zero Noise Extrapolation (ZNE), which are automatized within the \texttt{Qiskit} Runtime Estimator primitive \cite{estimator}.  
\subsection{Appendix B: Simulation implementation of QFIAE}

We have also implemented the QFIAE algorithm using two different simulation frameworks. \texttt{Pennylane}~\cite{pennylane} has been used for QNN implementation, whereas \texttt{Qibo}~\cite{qibo_paper} has been used for applying  IQAE to the Fourier series.

In particular, the linear Ansatz corresponding to each layer $\mathcal{L}_{LA}^{(l)}(\vec{x},\vec{\theta})$ for training the QNN to fit a $M$-dimensional function is the following \cite{Casas:2023ure}:
\begin{equation}
    \mathcal{L}_{LA}^{(l)}(\vec{x},\vec{\theta})=\prod_{i=1}^M\mathcal{S}(x_i)\mathcal{A}_i^{(l)}(\vec{\theta}_{l,i})~,
    \label{eq:linear_ansatz}
\end{equation}
where $\mathcal{S}(x_i)$ and $\mathcal{A}_i^{(l)}$ are chosen as:
\begin{equation}
    \mathcal{S}(x_i)=R_z(x_i)~, \quad \quad \mathcal{A}_i^{(l)}(\vec{\theta}_{l,i})=R_z(\theta_{l,i,1})R_y(\theta_{l,i,2})R_z(\theta_{l,i,3})~.
    \label{eq:SandA}
\end{equation}

Once the Fourier coefficients are obtained from the QNN, we implement the IQAE algorithm. To uphold the claimed quantum advantage provided by Grover's amplitude amplification, certain conditions must be fulfilled. First, the probability distribution of the functions to be integrated should be encodable into a shallow quantum circuit. In view of this requirement, we will use the distribution $p(x_i)=1/2^n$ generated by applying an $n$-dimensional Hadamard gate, denoted as $\mathcal{H}^{\otimes n}$, which corresponds to a quantum circuit of depth 1. The second condition is that the target function has to be encodable with a minimum number of quantum arithmetic operations. That will be achieved selecting the target function as a $\sin(x_i)^2$ to be integrated in $[x_{i,min},x_{i,max}]$. Then the integrals of the Fourier terms are obtained from the integral of the sine function.

\begin{figure}[th]
       \centering
       \begin{subfigure}[b]{0.49\textwidth}
       \centering
           \includegraphics[width=.99\linewidth]{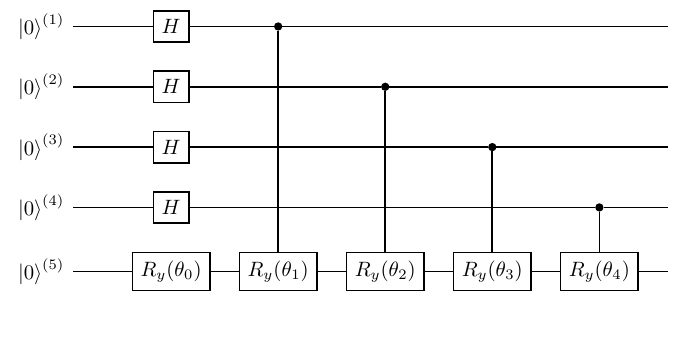}  
           \caption{Amplitude operator $\mathcal{A}$}
       \end{subfigure}
       \hfill
         \begin{subfigure}[b]{0.49\textwidth}
         \centering
           \includegraphics[width=.99\linewidth]{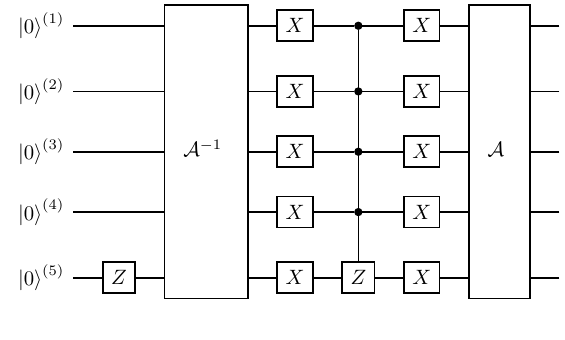} 
           \caption{Amplification operator $\mathcal{Q}$}
       \end{subfigure}
        \caption{Quantum circuits of the operators for the IQAE component of the QFIAE algorithm.
        }
        \label{fig:qaecircuits}
\end{figure}

Under these considerations and choosing $n_{qubits}=5$, the quantum circuits corresponding to the $\mathcal{A}$ and $\mathcal{Q}$ operators are shown in Fig.~\ref{fig:qaecircuits}.
Note that in Fig.~\ref{fig:qaecircuits}(a) the rotation angles encode the information about the limits of integration $x_{i,min}$ and $x_{i,max}$. In particular, they are defined as:
\beq
\theta_0=(x_{max} - x_{min}) / 2 ^ n + 2x_{min}, \quad \theta_i=2^{(i+ 1)} ( x_{max} - x_{min}) / 2 ^ n, \quad n=n_{qubits}.
\eeq
For more information, a tutorial on QFIAE implementation can be found at \cite{tutorial_qfiae}.

\subsection{Appendix C. Parameters employed for integrals implementation}

See tables \ref{table:paramsqnn} and \ref{table:parasiqae} below:

\begin{table}[h!]
    \centering
\begin{minipage}[b]{0.99\linewidth}\centering
    \begin{tabular}{ |p{1.65cm}|p{1.85cm}|p{1.85cm}|p{1.95cm} |p{1.95cm}|p{1.95cm}|p{1.95cm}| }
 \hline
 \centering  & \centering Tadpole (HW) & \centering Tadpole (SIM) &  \centering Bubble $m/p_0<0.5$ (SIM) & \centering Bubble $m/p_0>0.5$ (SIM) &  \centering Triangle $m/\sqrt{s}<0.5$ (SIM) & \centering Triangle $m/\sqrt{s}>0.5$ (SIM)\cr
 \hline
  \centering \textit{layers} & \centering 3 & \centering 5 &  \centering 10 &  \centering 20  &  \centering 10 &  \centering 10 \cr
 \hline
  \centering \textit{n\_{Fourier}} & \centering 5 & \centering 5 &  \centering 10 &  \centering 20 &  \centering 10\footnotemark &  \centering 10 \cr 
 \hline
    \centering \textit{step\_size} & \centering 0.100 & \centering 0.060 &  \centering 0.095 &  \centering 0.020 &  \centering \centering 0.065 &  \centering 0.045 \cr
\hline
 \centering \textit{max\_steps} & \centering 60 & \centering 300 &  \centering 450 &  \centering 200 &  \centering 450 &  \centering 400  \cr
\hline
 \centering \textit{data\_train} & \centering 15 & \centering 150 &  \centering 150 &  \centering 1500  &  \centering 3600 &  \centering 10000  \cr
\hline 
\centering \textit{shots} & \centering 500 & \centering - &  \centering - &  \centering - & \centering - &  \centering -  \cr
\hline
\centering \textit{$\delta$} & \centering - & \centering -&  \centering 21 &  \centering - & \centering 0.1 &  \centering - \cr 
\hline

\end{tabular}
\caption{Parameters for training the QNN.}
\label{table:paramsqnn}
\setcounter{mpfootnote}{0}
\footnotetext{Note that since it is a 2D function, the total number of Fourier coefficients is \textit{n\_{Fourier}}$^2=100$.  }
\end{minipage}
\end{table}

\begin{table}[h!]
    \centering

    \begin{tabular}{ |p{1.65cm}|p{1.95cm}|p{1.85cm} |p{1.85cm}| }
 \hline
 \centering  & \centering Tadpole \\ (HW \& SIM) &  \centering Bubble (SIM) & \centering Triangle (SIM) \cr
 \hline
  \centering $\varepsilon$ & \centering 0.01 &  \centering 0.001 &  \centering 0.001 \cr 
 \hline
  \centering $\alpha$ & \centering 0.05 &  \centering 0.05 &  \centering 0.05 \cr 
 \hline
    \centering \textit{shots} & \centering 100000 &  \centering 1000 &  \centering 10000  \cr
\hline

\end{tabular}
\caption{Parameters for the IQAE.}
\label{table:parasiqae}
\end{table}

\end{document}